\documentclass{article}
\usepackage{frascatiphys}
\usepackage{graphicx}
\usepackage{amsmath}
\usepackage{amsfonts}
\usepackage{bbold}
\newcommand{\beq}{\begin{equation}}
\newcommand{\eeq}{\end{equation}}
\newcommand{\beqa}{\begin{eqnarray}}
\newcommand{\eeqa}{\end{eqnarray}}
\def\lsim{\raise0.3ex\hbox{$<$\kern-0.75em\raise-1.1ex\hbox{$\sim$}}}
\def\gsim{\raise0.3ex\hbox{$>$\kern-0.75em\raise-1.1ex\hbox{$\sim$}}}

\def\0{{\bf 0}}

\def\kk{{\kappa}}
\def\pp{{\hat{p}}}

\begin{document}
\title{Heavy-flavour in relativistic nuclear collisions: recent developments}

\author{
Andrea Beraudo        \\
{\em INFN - Sezione di Torino, Via Pietro Giuria 1 - 10125 Torino}}
\maketitle
\baselineskip=10pt
\begin{abstract}
 Transport calculations represent the major tool to simulate the modifications induced by the presence of a hot-deconfined medium on the production of heavy-flavour particles in high-energy nuclear collisions. After a brief description of the approach  and of the major achievements in its phenomenological applications we discuss some recent developments. In particular we focus on observables arising from event-by-event fluctuations in the distribution of deposited energy (odd flow harmonics, event-shape-engineering) and from the tilting of the initial geometry with respect to the beam axis (directed flow), with a possible role played by the strong magnetic field generated by the spectator nucleons.
\end{abstract}

\section{Introduction}
Heavy-flavour particles play a peculiar role in probing the hot-deconfined matter produced in relativistic heavy-ion collisions. Soft observables (light hadrons of low transverse-momentum) provide information on the collective behaviour of the medium formed after the collision; they are nicely described by hydrodynamic calculations, assuming as a working hypothesis to deal with a system close to local thermal equilibrium. The suppression of the production of jets and high-$p_T$ particles tells us that a quite opaque medium is formed in the collisions: its description requires to model the energy-loss of high-energy partons in the hot plasma.
Heavy-flavour particles, arising from the hadronization of heavy quarks produced in initial hard events and having crossed the fireball during its whole evolution, require to employ a more general tool, allowing one to model their asymptotic approach to local thermal equilibrium with the medium: such a tool is represented by transport calculations, which we are going to briefly describe.
Actually, since the fireball undergoes a rapid expansion and has a finite lifetime, one does not expect charm and beauty quarks to reach full kinetic equilibrium with the medium: this fact, however, has the potential to provide an estimate of the value of the transport coefficients of the medium, for which otherwise one would get just a lower/upper bound.

\section{Transport calculations}
\begin{figure}[!ht]
\begin{center}
\includegraphics[clip,width=0.49\textwidth]{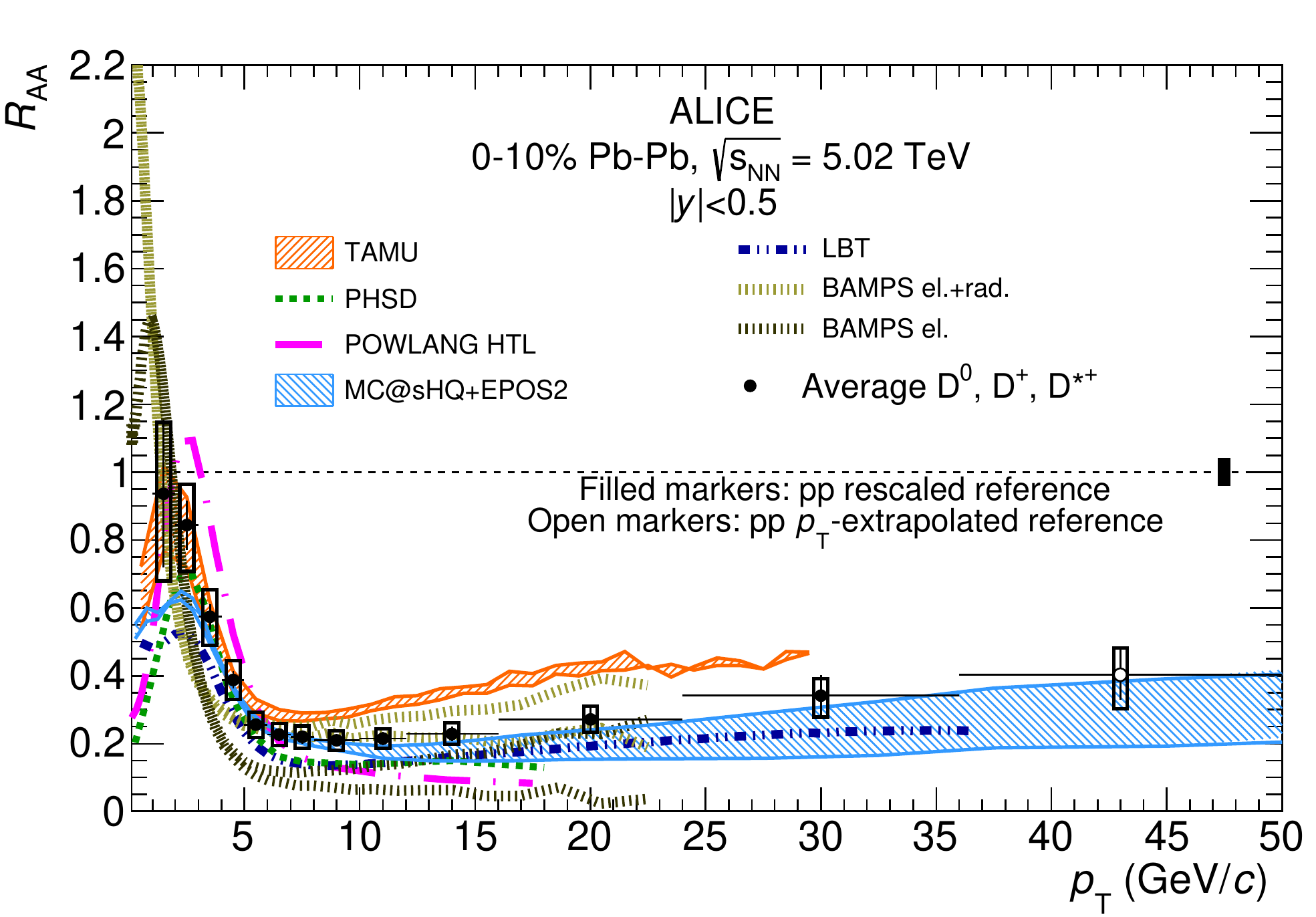}
\includegraphics[clip,width=0.49\textwidth]{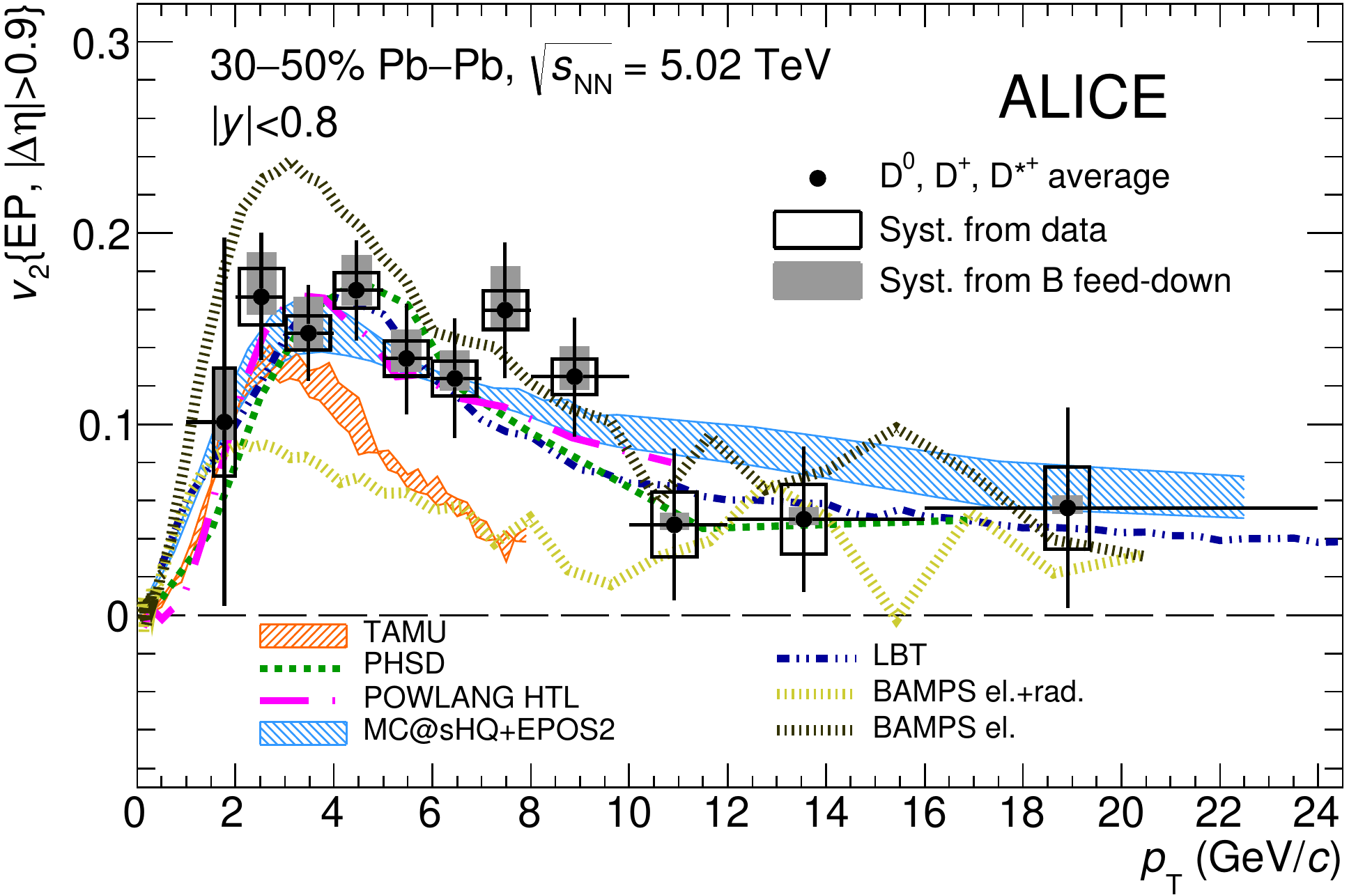}
\caption{Recent results by the ALICE collaboration for the average $D$-meson $R_{\rm AA}$ and $v_2$~\cite{Acharya:2018hre} compared to the predictions of various transport calculations.}\label{fig:data-vs-models} 
\end{center}
\end{figure}

The starting point of any transport calculation is the relativistic Boltzmann equation. Actually, in most numerical implementation, the latter is approximated as a more tractable Langevin equation, assuming that the heavy-quark interaction with the medium is dominated by multiple uncorrelated soft scatterings. One has then
\beq
{\Delta \vec{p}}/{\Delta t}=-{\eta_D(p)\vec{p}}+{\vec\xi(t)}.\label{eq:Langevin}
\eeq
Eq.~(\ref{eq:Langevin}) provides a recipe to update the heavy quark momentum in the time-step $\Delta t$ through the sum of a deterministic friction force and a random noise term specified by its temporal correlator
\beq
\langle\xi^i(\vec p_t)\xi^j(\vec p_{t'})\rangle\!=\!{b^{ij}(\vec p_t)}{\delta_{tt'}}/{\Delta t}\quad{\rm with}\quad{b^{ij}(\vec p)}\!\equiv\!{\kk_\|(p)}\pp^i\pp^j+{\kk_\perp(p)}(\delta^{ij}\!-\!\pp^i\pp^j).
\eeq
Following the heavy-quark dynamics in the medium requires then the knowledge of three transport coefficients representing the transverse/longitudinal ($\kappa_{\perp/\|}$) momentum broadening and the drag ($\eta_D$) received from the medium. Actually, the above coefficients are not independent, but are related by the Einstein fluctuation-dissipation relation, which ensures the asymptotic approach of the heavy quarks to thermal equilibrium.

Various transport calculations applied to heavy-flavour production in nuclear collisions can be found in the literature, essentially differing in the choice of transport coefficients to insert into Eq.~(\ref{eq:Langevin}). The challenge for the above models is to consistently reproduce various experimental observables, like the momentum and angular distributions of the produced heavy-flavour hadrons. The latter display sizable modifications with respect to proton-proton collisions. In particular, important medium effects are captured by two quantities, the nuclear modification factor $R_{\rm AA}$ and the elliptic-flow coefficient $v_2$, defined as
\beq
R_{\rm AA}\equiv\frac{\left(dN/dp_T\right)_{\rm AA}}{\langle N_{\rm coll}\rangle\left(dN/dp_T\right)_{\rm AA}}\quad{\rm and}\quad v_2\equiv\langle\cos[2(\phi-\Psi_{\rm RP})]\rangle.
\eeq
The $R_{\rm AA}$ is the ratio of the momentum distribution measured in A-A and p-p collisions normalized to the average number of independent binary nucleon-nucleon collisions in a A-A event. Deviations from unity signal the presence of medium effects: at high $p_T$ one gets $R_{\rm AA}<1$, due to the energy-loss of charm and beauty quarks; instead, its rise at low-moderate $p_T$ may come from the collective radial flow of the fireball, boosting particles from very low to higher transverse momenta.
The $v_2$ coefficient quantifies the azimuthal anisotropy of the angular distribution of final particles and is interpreted as arising from the elliptic asymmetry of the initial condition in non-central collisions, with the larger pressure gradients along the reaction plane $\psi_{\rm RP}$ giving rise to a larger acceleration of the fluid along this direction.
The challenge for the models is to consistently reproduce these and other observables. 
A snapshot of the results of different transport calculations~\cite{Beraudo:2014boa,He:2014cla,Cao:2017hhk,Nahrgang:2013xaa} compared to ALICE data~\cite{Acharya:2018hre} is given in Fig.~\ref{fig:data-vs-models}. Notice that, in all cases, in order to reproduce the experimental data, it is important to include the possibility for heavy quarks to hadronize via recombination with the light thermal partons from the medium.

\section{Recent developments}
\begin{figure}[!ht]
\begin{center}
\includegraphics[clip,width=0.49\textwidth]{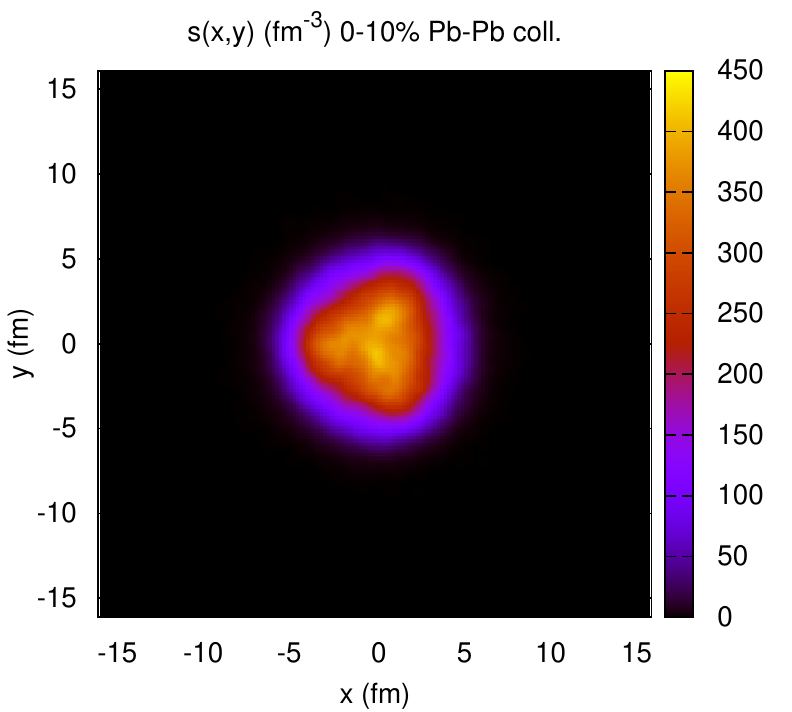}
\includegraphics[clip,width=0.49\textwidth]{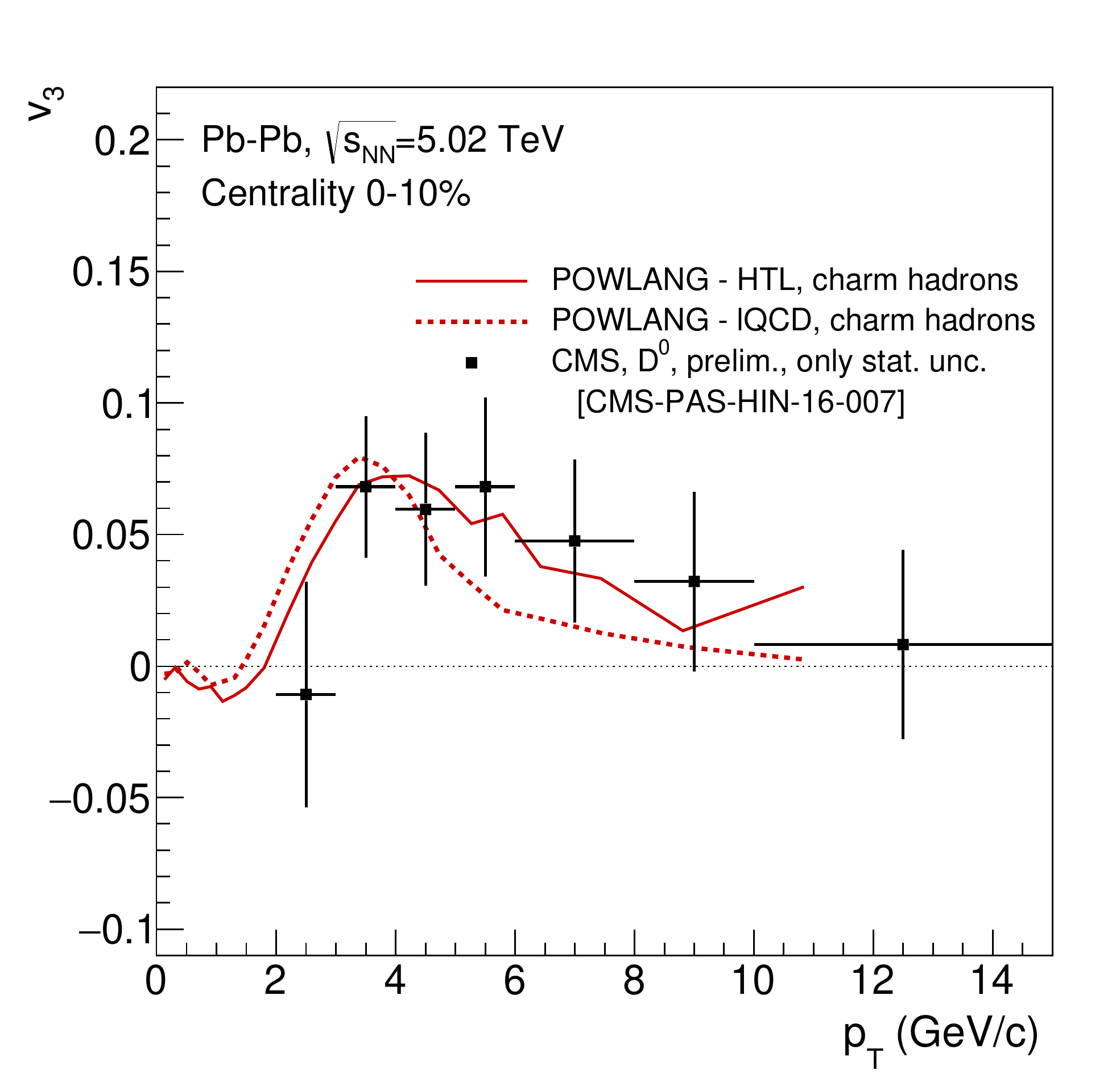}
\caption{Initial condition for a central Pb-Pb collision at the LHC displaying a triangular eccentricity (left panel) and the resulting $v_3$ coefficient (right panel) of the azimuthal distribution of $D$-mesons for different choices of the transport coefficients~\cite{Beraudo:2017gxw}.}\label{fig:e3-vs-v3} 
\end{center}
\end{figure}
The finite impact parameter of a nucleus-nucleus collision leads, on average, to an elliptic deformation of the produced fireball. Pressure gradients map this initial geometric asymmetry into a final momentum anisotropy of the particles decoupling from the medium, giving rise to the elliptic flow $v_2$ shown for instance in the right panel of Fig.~\ref{fig:data-vs-models}. However, event-by-event fluctuations (e.g. in the nucleon positions) can give rise to more complicated initial geometries, quantified by higher order eccentricity coefficients
\beq
\epsilon_{m}=\frac{\sqrt{\{r_\perp^2\cos(m\phi)\}^2+\{r_\perp^2\sin(m\phi)\}^2}}{\{r_\perp^2\}}
\eeq
which lead to higher harmonics in the final hadron distributions $v_m\equiv\langle\cos[m(\phi-\Psi_m)]\rangle$.
In Fig.~\ref{fig:e3-vs-v3} we show the result of a one-shot hydro+transport simulation starting from an average initial condition with a triangular deformation referring to the 0-10\% most central Pb-Pb collisions at $\sqrt{s_{\rm NN}}\!=\!5.02$ TeV. The final $D$-meson angular distribution is then characterized by a non-vanishing triangular flow $v_3$, as shown in the right panel of the figure where we compare the results of our transport simulations~\cite{Beraudo:2017gxw} to CMS data~\cite{Sirunyan:2017plt}.

\begin{figure}[!ht]
\begin{center}
\includegraphics[clip,width=0.49\textwidth]{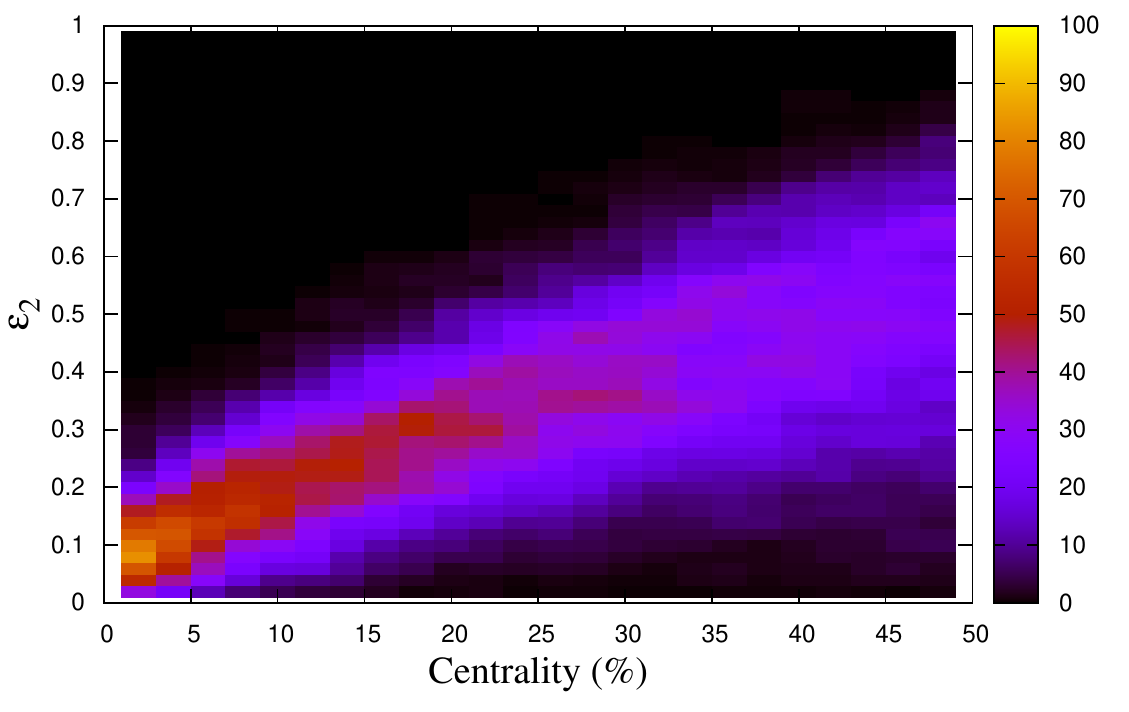}
\includegraphics[clip,width=0.49\textwidth]{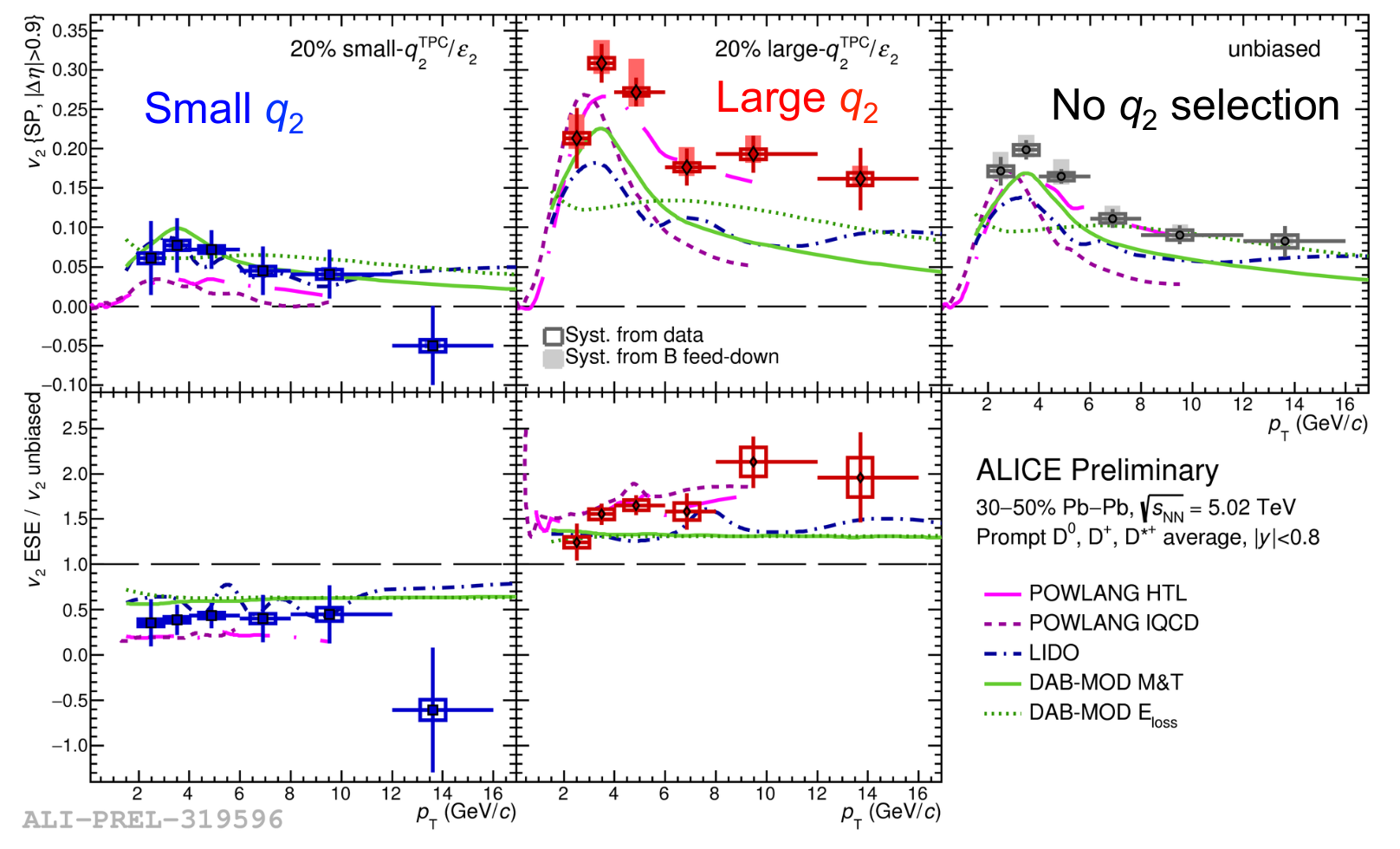}
\caption{Left panel: (elliptic) eccentricity-vs-centrality correlations in Pb-Pb collisions at the LHC from event-by-event Glauber-MC simulations of the initial condition~\cite{Beraudo:2018tpr}. Right panel: $D$-meson elliptic flow after selecting events, within a given centrality class, with high/low eccentricity. ALICE data~\cite{Acharya:2018bxo} are compared to the predictions of various transport models.}\label{fig:ese-1} 
\end{center}
\end{figure}

\begin{figure}[!ht]
\begin{center}
\includegraphics[clip,width=0.9\textwidth]{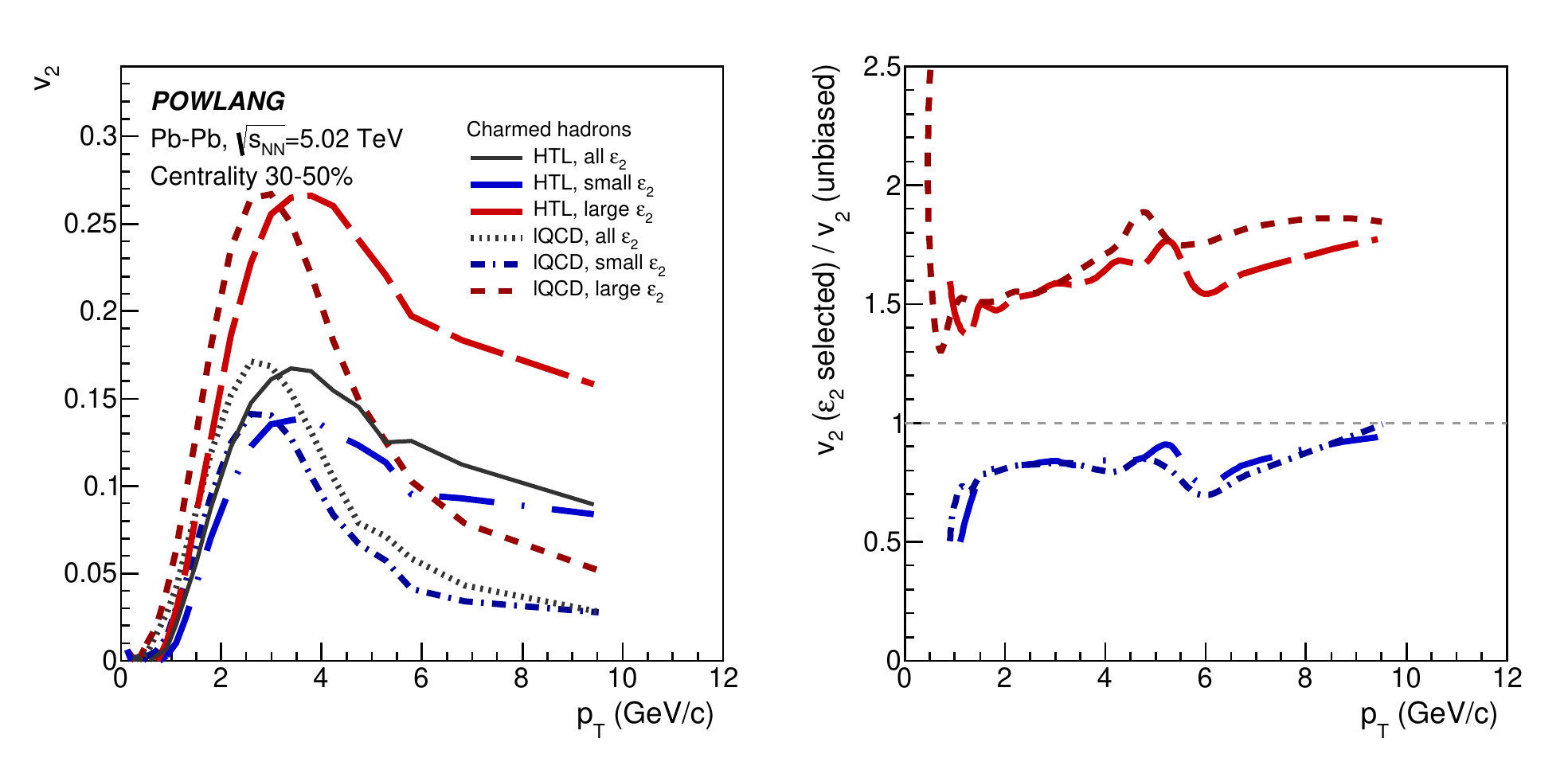}
\caption{$D$-meson elliptic flow in the 30-50\% centrality class for events with high/low initial eccentricity compared to an unbiased selection of events~\cite{Beraudo:2018tpr}. The ratio to the unbiased result (right panel) turns out to be independent of the choice of the transport coefficients.}\label{fig:ese-2} 
\end{center}
\end{figure}
Furthermore, due to event-by-event fluctuations, events belonging the same centrality class -- usually identified by some estimator like the number of binary nucleon-nucleon collisions (in numerical simulations) or the multiplicity of produced particles (in actual experiments) -- can be characterized by quite different initial eccentricities, as shown in the left panel of Fig.~\ref{fig:ese-1}. It is then of interest to study, for a given centrality, the elliptic (or triangular) flow of subsample of events of high/low eccentricity, comparing the result to the unbiased case. This technique, known as \emph{event-shape-engineering}, was first introduced for light hadrons~\cite{Adam:2015eta} and later applied also to study of the flow of $D$-mesons~\cite{Acharya:2018bxo}. Results obtained by the ALICE collaboration, compared to various transport calculations, are displayed in the right panel of Fig.~\ref{fig:ese-1}. In Fig.~\ref{fig:ese-2} we show the results of the transport model of Ref.~\cite{Beraudo:2018tpr} for the $D$-meson elliptic flow in Pb-Pb collisions, in the 30-50\% centrality class, for the 0-20\% highest and 0-60\% lowest-eccentricity subsamples. Notice how the ratio to the unbiased result does not depend on the choice of the transport coefficients, reflecting only the initial geometry. This holds quite generally also for beauty hadrons and for different centrality classes.

\begin{figure}[!ht]
\begin{center}
  \includegraphics[clip,height=4.5cm]{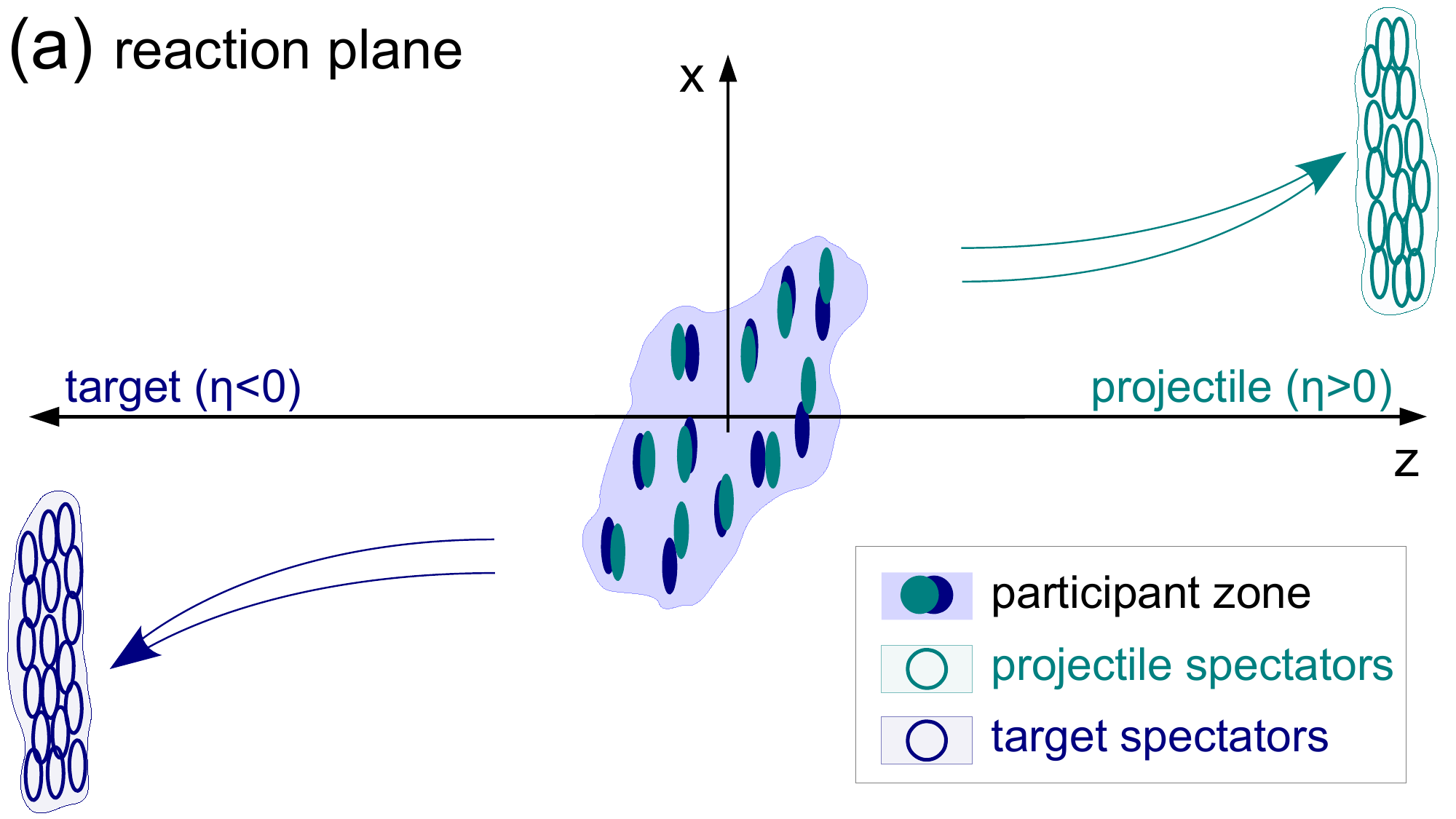}
  \includegraphics[clip,height=4.5cm]{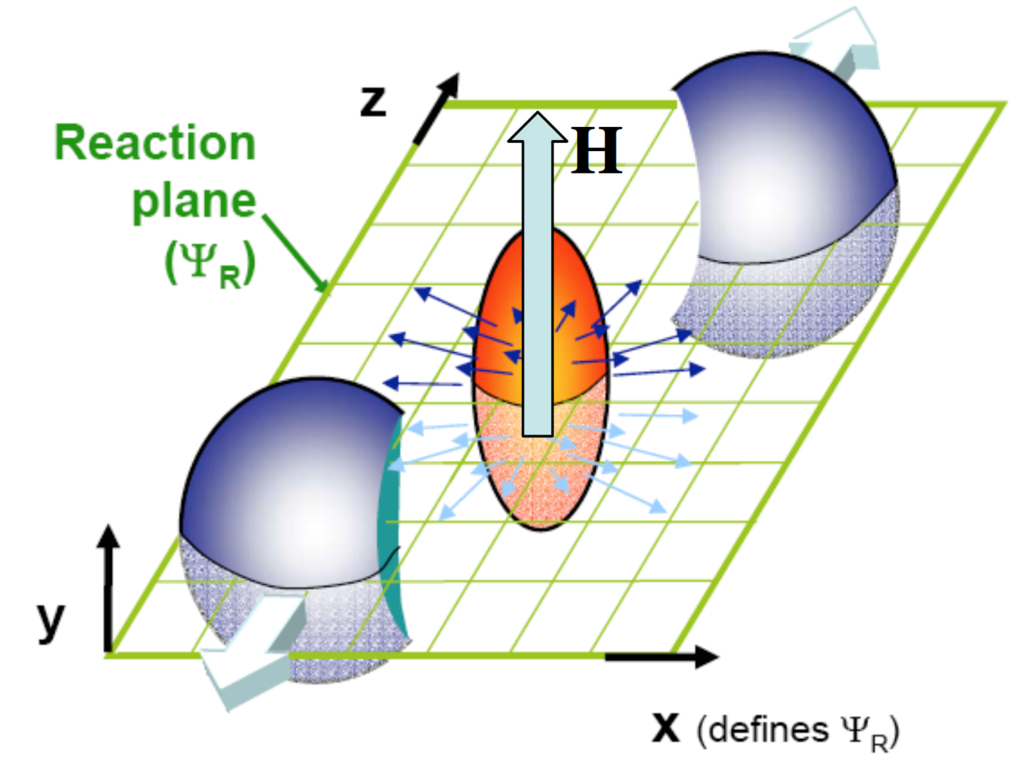}  
\caption{Schematic picture of a non-central heavy-ion collision, showing the tilting of the initial condition with respect to the beam axis (left panel) and the magnetic field orthogonal to the reaction plane arising from the spectator nucleons.}\label{fig:v1-exp} 
\end{center}
\end{figure}

\begin{figure}[!ht]
\begin{center}
  \includegraphics[clip,height=5cm]{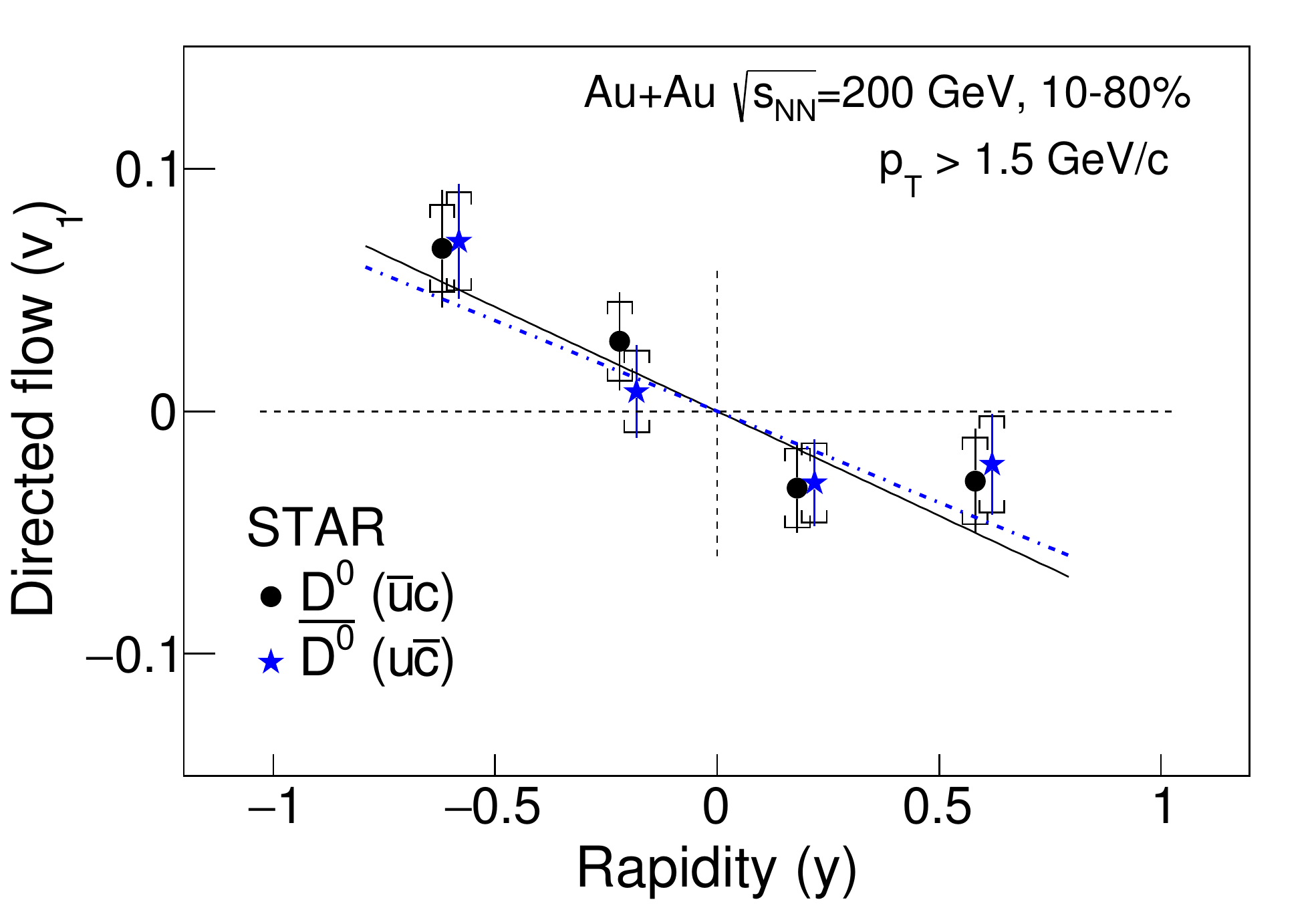}
  \includegraphics[clip,height=5cm]{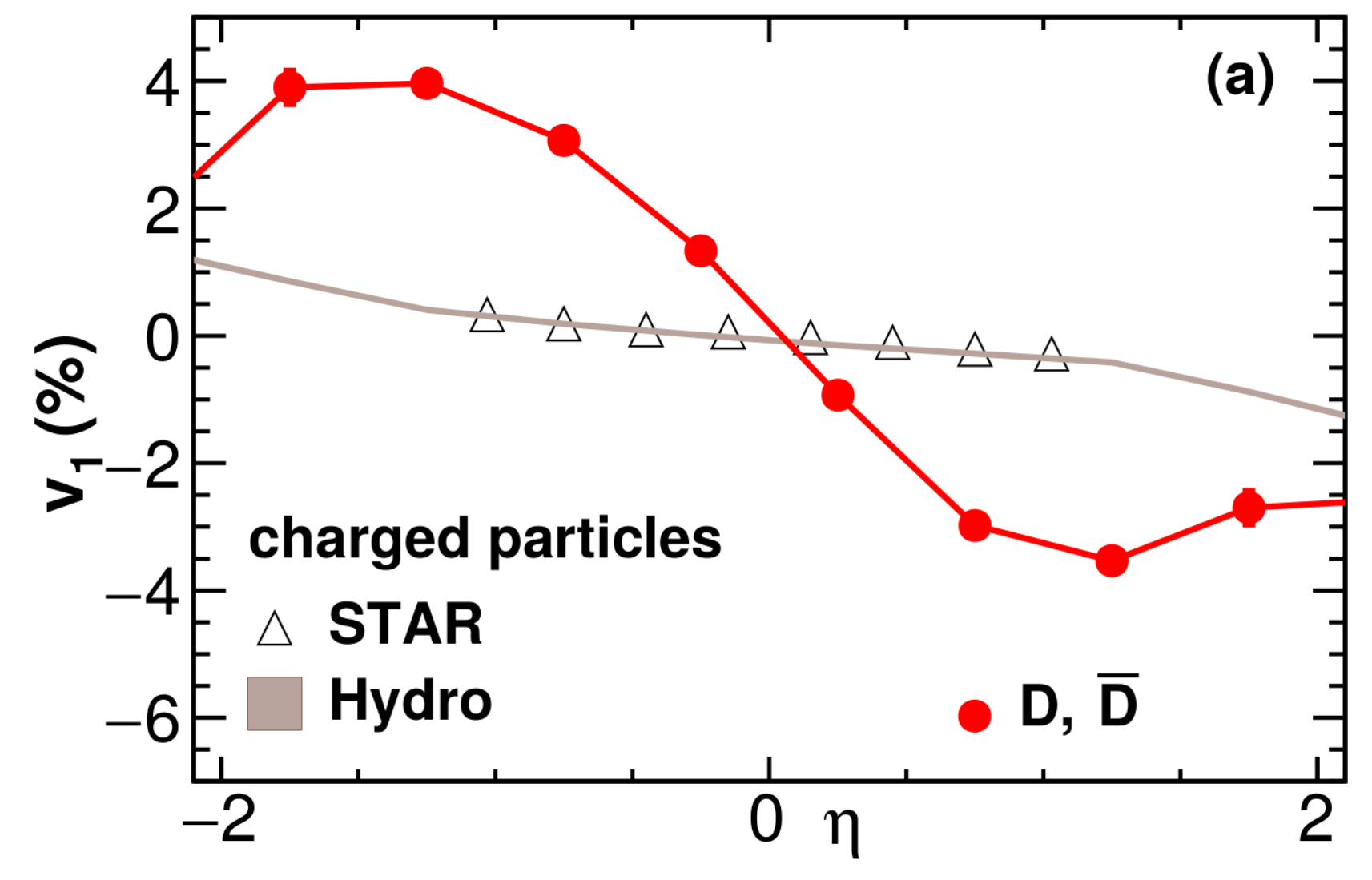}  
\caption{The directed flow $v_1$ of $D^0/\overline D^0$ mesons in Au-Au collisions at $\sqrt{s_{\rm NN}}\!=\!200$ GeV as a function of rapidity measured by the STAR collaboration~\cite{Adam:2019wnk} and predicted by a transport calculation~\cite{Chatterjee:2017ahy}.}\label{fig:v1-th} 
\end{center}
\end{figure}
Recently, a strong interest is growing also for the study of the directed flow $v_1\!\equiv\!\langle\cos(\phi\!-\!\Psi_{\rm RP})\rangle$. Since participant nucleons of the colliding nuclei tend to deposit more energy along their direction of motion, in non-central collisions the fireball is characterized by an initial tilted geometry (see left panel of Fig.~\ref{fig:v1-exp}) and by a sizable orbital angular momentum (of order 1000$\hbar$) and vorticity~\cite{Becattini:2015ska}. Experimentally, this can give rise to a negative/positive directed flow $v_1$ of charged hadrons at forward/backward rapidity and possibly to other effects like the polarization of $\Lambda$ hyperons~\cite{STAR:2017ckg}. Interestingly, one expects a stronger $v_1$ signal (quite small for light hadrons) in the case of charmed particles. On top of the directed flow of the background medium an important contribution to the final signal arises in fact from the mismatch between the tilted geometry of the medium and the initial position of the $c\overline c$ pairs, symmetrically distributed around the beam axis. Predictions for the $D$-meson $v_1$ of the transport calculation of Ref.~\cite{Chatterjee:2017ahy} are shown in the right panel of Fig.~\ref{fig:v1-th}. The comparison with experimental data should allow one to probe the three-dimensional distribution of matter in heavy-ion collisions. Recently, some authors~\cite{Das:2016cwd,Chatterjee:2018lsx} have also proposed that the difference between the $v_1$ of $D^0$ and $\overline{D^0}$ mesons can be a unique probe of the huge electromagnetic fields present in the fireball during the deconfined phase; however, current experimental data~\cite{Adam:2019wnk} do not allow yet to draw firm conclusions.

\section{Conclusions and perspectives}
The study of heavy-flavour observables in nucleus-nucleus collisions has the potential to provide information on charm and beauty transport coefficients in the quark-gluon plasma, describing their spatial diffusion and their approach to kinetic equilibrium. For the moment several theoretical calculations allow one to get a quite satisfactory description of the data, leading then to a substantial theoretical uncertainty on the value of the above coefficients. In-medium hadronization plays a role at least as important as the one of the transport in the partonic phase; furthermore, so far, one has experimental access to a kinematic window where the heavy-quark dynamics can not be captured simply by one non-relativistic transport coefficient. We expect the situation to improve as soon as measurements of beauty at low transverse-momentum will get available.

Besides the transport coefficients, the study of heavy flavour observables provides information on other non-trivial feature nuclear collisions, like the event-by-event fluctuations and the initial three-dimensional distribution of deposited energy and, possibly, on the effect of the initial huge magnetic field arising from the spectator nucleons. In this connection, we tried to give an overview of the most recent theoretical studies and experimental analysis.

\bibliography{trento}

  \end{document}